\documentclass[pre,twocolumn]{revtex4}
\usepackage[english]{babel}
\usepackage{amsmath}
\usepackage{amssymb}
\usepackage{epsfig}

\begin{document}
\title{An Optical Analog for a Rotating Binary Bose--Einstein Condensate}
\author{Victor P. Ruban}
\email{ruban@itp.ac.ru}
\affiliation{Landau Institute for Theoretical Physics RAS,
Chernogolovka, Moscow region, 142432 Russia}

\date{\today}

\begin{abstract}
Coupled nonlinear Schr\"odinger equations for paraxial optics with
two circular polarizations of light in a defocusing Kerr medium with
anomalous dispersion coincide in form with the Gross--Pitaevskii
equations for a binary Bose--Einstein condensate (BEC) of cold 
atoms in the phase separation regime. In this case, the helical 
symmetry of an optical waveguide corresponds to rotation of the 
transverse potential confining the BEC. The “centrifugal force” 
considerably affects the propagation of a light wave in such a system.
Numerical experiments for a waveguide with an elliptical cross 
sections have revealed characteristic structures consisting of 
quantized vortices and domain walls between two polarizations, 
which have not been observed earlier in optics.
\end{abstract}

\maketitle
%%%%%%%%%%%%

\section{Introduction}

It is known that the propagation of a quasi-monochromatic light
wave in an optical medium with Kerr nonlinearity can be approximately
described by a nonlinear Schr\"odinger equation (NLSE) 
(see, for example, [1--4] and the literature cited therein). 
In this case, a wave in a defocusing medium with anomalous group
velocity dispersion resembles a rarefied Bose--Einstein condensate
(BEC) of cold repelling atoms, which is characterized by soft
topological excitations in the form of quantum vortices [5]. 
An optical wave may have two circular polarizations, and then 
light is described by two coupled NLSEs [6] analogously to 
a binary BEC [7--13]. In a binary system, apart from dark solitons
and quantized vortices in each of the two components, domain walls
separating the regions with right- and left-hand circular polarizations
are also possible [14--22]. A domain wall exhibits an effective surface
tension [10, 23], which strongly affects the dynamics of the regions. 
In the theory of BECs, the phase separation and phenomena accompanying
it have been investigated quite thoroughly (see, for example, [24--48]
and the literature cited therein). As regards nonlinear optics, 
essentially three-dimensional (3D) binary optical structures have been
revealed recently in [21, 22].

This study is devoted to yet insufficiently investigated nonlinear 
quasi-monochromatic light beam with two polarizations in a wide 
waveguide with helical symmetry, in which the spatial inhomogeneity 
of refractive index $n=n_0+\tilde n(x,y,\zeta)$ of the medium at carrier 
frequency $\omega_0$ depends on two combinations of dimensionless variables
$$
\tilde x= x \cos\Omega \zeta +y \sin\Omega \zeta,\qquad
\tilde y= y \cos\Omega \zeta -x \sin\Omega \zeta.
$$
Depending on the system under consideration, parameter $\Omega$ 
determines either the helical waveguide pitch or the period of 
rotation of the atomic trap. The small deviation has form
$$
\tilde n \propto -U(\tilde x,\tilde y),
$$
where $U$ is a certain two-dimensional (2D) well. It should be noted
at the very outset that instead of time $t$, the role of the evolution
variable is played by distance $\zeta$ along the waveguide axis, 
while the role of the third “spatial” coordinate is played by “delayed”
time
$$
\tau=t-\zeta/v_{\rm gr}.
$$
To compare with BECs, one must consider $t$ instead of $\zeta$ and $z$
instead of $\tau$. This type of waveguides is interesting because in
the physics of cold gases, it corresponds to the class of uniformly
rotating external potentials
\begin{equation}
V(x,y,t)=U(x\cos\Omega t+y \sin\Omega t, y\cos\Omega t -x \sin\Omega t).
\end{equation}
From the point of view of applications, such potential wells for atoms,
which are infinitely large in coordinate $z$, are hardly realistic.
However, in optics, such wells indicate just a possible infinitely 
large length of an optical beam in variable $\tau$, which is a quite
admissible approximation. The rotation of a BEC is known to lead to 
the emergence of quantized vortices with various configurations in it.
The presence of two components additionally supplements the pattern
with domain walls [34--37]. Accordingly, analogous regimes must take 
place in optics also. This study is aimed at the observation of 
previously unknown coherent binary structures in optical experiments.

\section{Equations and numerical method}
    
We consider a 3D optically transparent dielectric medium with the 
defocusing Kerr nonlinearity. The medium is weakly inhomogeneous 
in space. The background permittivity is given by function of frequency 
$\varepsilon(\omega)$ so that the corresponding dispersion
relation has form
$$
k(\omega)=\sqrt{\varepsilon(\omega)}\omega/c.
$$
We are interested in the range of anomalous dispersion of group
velocity (where $k''(\omega)<0$). As a rule, such a range is 
near the low-frequency edge of the transparency window (in actual 
substances, this is often the infrared spectral region; see, 
for example, [49, 50]).

The equations for slow complex amplitudes $A_{1,2}(x,y,\tau,\zeta)$ 
of the left- and right-hand polarizations of a light wave are well known
[6, 14--21]. For this reason, we will only remind here in brief the main
idea of derivation of these equations. The starting point is the 
corollary to Maxwell equations
\begin{equation}
\mbox{curl}\,\mbox{curl}\,{\textbf{\em E}}=
-\frac{1}{c^2}\frac{\partial^2 {\textbf{\em D}}}{\partial t^2}.
\label{wave_eq}
\end{equation}
We introduce slow complex envelopes ${\bf E}$ and ${\bf D}$ by substitutions
\begin{eqnarray}
{\textbf{\em E}}&=&\mbox{Re}[{\bf E}\exp(ik_0\zeta-i\omega_0 t)],\nonumber\\
{\textbf{\em D}}&=&\mbox{Re}[{\bf D}\exp(ik_0\zeta-i\omega_0 t)],\nonumber
\end{eqnarray}
where $k_0=k(\omega_0)$ is the carrier wavenumber. Further, we must
substitute into Eqs.(2) the weakly nonlinear material dependence
\begin{eqnarray}
{\bf D} &\approx& \int\varepsilon(\omega_0+\tilde\omega)
{\bf E}_{\tilde\omega} e^{-i\tilde\omega t}\frac{d\tilde\omega}{2\pi}
\nonumber\\
&+&\tilde\varepsilon(x,y,\zeta){\bf E}
+\alpha(\omega_0)|{\bf E}|^2  {\bf E}
+\beta(\omega_0)({\bf E}\cdot{\bf E}) {\bf E}^*,
\end{eqnarray}
with negative functions $\alpha$ and $\beta$ in the defocusing case. 
In the course of transformations, there appears an equation of form
$
\hat L {\bf E}=s\{{\bf E}\},
$
with linear operator
$$
\hat L =(k_0-i\partial_\zeta)^2-[k(\omega_0+i\partial_t)]^2
$$
and a small right-hand side that includes the transverse Laplacian,
the nonlinearity, and the spatial inhomogeneity. In the main order in
smallness $s$, we can set
\begin{equation}
\hat L\approx 2k_0(-i\partial_\zeta-ik_0'\partial_t+k_0''\partial_t^2/2).
\end{equation}
At the end of derivation, to pass to scalar functions $A_{1,2}$, we perform
the substitution
\begin{equation}
{\bf E}\approx \big[({\bf e}_x+i{\bf e}_y) A_1 
    + ({\bf e}_x-i{\bf e}_y) A_2 \big]/\sqrt{2}.
\end{equation}

It is convenient to take for the scale of transverse coordinates a certain
large parameter $R_0$ (on the order of several tens of wavelengths,
i.e., up to a hundred of micrometers). We will measure longitudinal 
coordinate $\zeta$ in the units of $k_0 R_0^2$ (several centimeters) and 
variable $\tau$, in the units of $R_0\sqrt{k_0 |k_0''|}$, 
while the electric field will be measured in the units of 
$\sqrt{2\varepsilon(\omega_0)/|\alpha(\omega_0)|}/(k_0 R_0)$. 
Then the confining potential is defined as
$$
V=-k_0^2 R_0^2\tilde\varepsilon/2\varepsilon(\omega_0).
$$
In these dimensionless variables, coupled nonlinear Schr\"odinger 
equations have form
\begin{eqnarray}
i\frac{\partial A_{1,2}}{\partial \zeta}&=&\Big[-\frac{1}{2}\Delta +V(x,y,\zeta)
\nonumber\\
&&\qquad +|A_{1,2}|^2+ g_{12}|A_{2,1}|^2\Big]A_{1,2},
\label{A_12_eqs}
\end{eqnarray}
where
$
\Delta=\partial_x^2+\partial_y^2+\partial_\tau^2
$
is the 3D Laplace operator in the “coordinate” space  $(x,y,\tau)$.
The cross-phase modulation parameter
$$
g_{12}=1+2\beta(\omega_0)/\alpha(\omega_0)
$$
is typically equal approximately to 2, which corresponds to the 
phase separation regime.

In contrast to recent publication [21], where “nonrotating” parabolic
potentials $(x^2+\kappa^2y^2)/2$ have been considered, we concentrate
our attention on helical waveguides with a flat bottom and sharp walls.
Such waveguides are easier for experimental implementation. However, 
smooth potentials are more convenient for numerical simulation;
for this reason, we will approximate the corresponding rectangular well
by expression
\begin{equation}
U= C[1-\exp(-[(\tilde x^2+\kappa^2\tilde y^2)/36]^5)],
\end{equation}
with large parameter $C\sim 50$ and with anisotropy $\kappa>1$
(in all calculations, we set $\kappa^2=1.5$). As a result of such 
a choice, the effective waveguide diameter is approximately 10 (up to 1 mm).
On such a scale, several hundreds of light wavelengths can be fitted. 
At “rotational frequency” $\Omega\sim 1$, about ten quantized vortices 
fit into the cross-sectional area, and the width of their cores depends 
on the wave intensity as $\xi\sim 1/\sqrt{I}$. Domain wall width $w$ is 
of the same order of magnitude. As will be seen further from numerical
results, the most interesting values are $I\lesssim 10$.

It is important that Eqs. (6) form the Hamilton system
$$
i \partial A_{1,2}/\partial \zeta=\delta{\cal H}/\delta A^*_{1,2}.
$$
The corresponding nonautonomous Hamiltonian is
\begin{eqnarray}
{\cal H}&=&\frac{1}{2}\int(|\nabla A_1|^2+|\nabla A_2|^2)dx dy d\tau \nonumber\\
&+&\int V(x,y,\zeta)(|A_1|^2+|A_2|^2)dx dy d\tau \nonumber\\
&+&\frac{1}{2}\int(|A_1|^4+|A_2|^4+2g_{12}|A_1|^2|A_2|^2)dx dy d\tau.
\end{eqnarray}
This functional is not conserved in the course of evolution. However,
since we have an autonomous system in the rotating coordinate system,
the integral of motion is the functional
\begin{equation}
{\cal H}_\Omega={\cal H}
-\Omega\int [A^{\dag} (iy\partial_x-ix\partial_y)A]dx dy d\tau,
\end{equation}
where $A=(A_1,A_2)^T$ is the two-component column. In addition,
the integrals of intensities 
$$
N_{1,2}=\int |A_{1,2}|^2 dx dy d\tau
$$
are also conserved.

For numerical simulation of Eqs. (6), we have used  the standard 
split-step Fourier method of the second order of accuracy in evolution
variable $\zeta$ in the initial (nonrotating) coordinate system. 
The computational domain in variables $x,y,\tau$ had the shape 
of a cube with side $6\pi$ with periodic boundary conditions. However, 
since the potential well is quite deep, functions $A_{1,2}$ decrease 
rapidly almost to zero in the transverse directions, so that the effect
of transverse boundaries is negligibly small.

The accuracy of calculations was monitored by the conservation of the 
integrals of motion to 3--6 decimal places on interval $0< \zeta < 500$ 
(it corresponds to the length of several tens of meters).

For the numerical preparation of the initial state with as small as
possible excitation of hard degrees of freedom, the imaginary-time 
propagation method was employed in a uniformly rotating coordinate system.
This method corresponds to the purely gradient dissipative dynamics
$$
-\partial A_{1,2}/\partial \eta=\delta\tilde{\cal H}/\delta A^*_{1,2},
$$
where modified Hamiltonian
$$
\tilde{\cal H}={\cal H}_\Omega-\mu(N_1+N_2)
$$
does not contain an explicit dependence on the quasi-time variable $\eta$.
Parameter $\mu$ (chemical potential) determines the characteristic values
of intensities
$$
I_{1,2}=|A_{1,2}|^2 \sim \mu.
$$
The interval for variable $\eta$ was several tens so that all hard degrees
of freedom were effectively suppressed and the system was in a slow dynamic
regime close to the minimum of $\tilde{\cal H}$.

\begin{figure}
\begin{center} 
(a)\epsfig{file=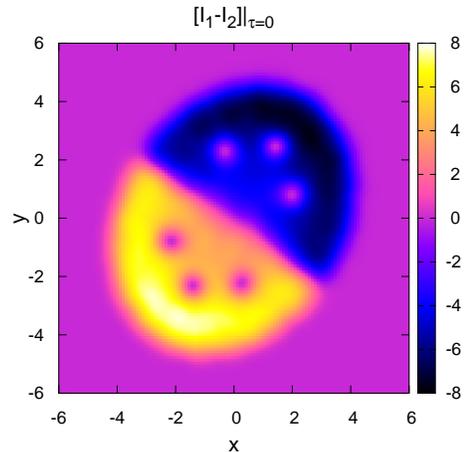, width=75mm}\\
(b)\epsfig{file=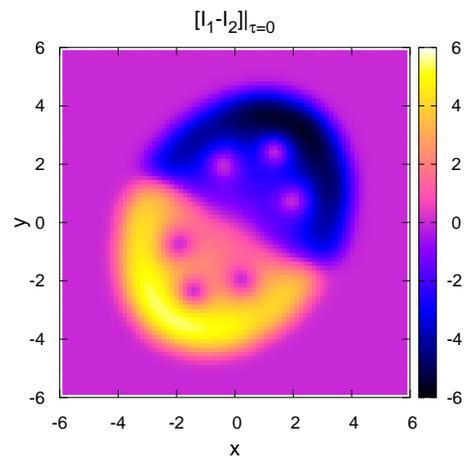, width=75mm}\\
(c)\epsfig{file=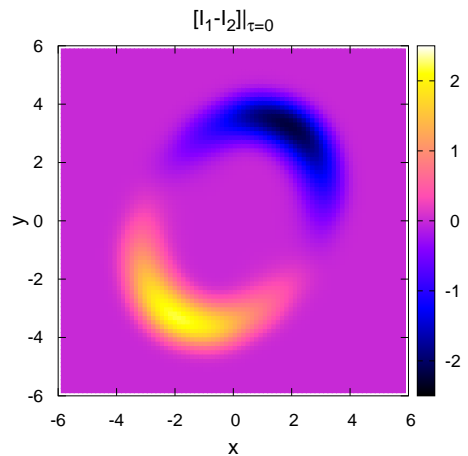, width=75mm}
\end{center}
\caption{ Quasi-two-dimensional configurations for $\zeta=500$, which
correspond to rotational frequency $\Omega=0.8$. The difference in
the local intensities in cross section $\tau=0$ is demonstrated for
three values of chemical potential: (a) $\mu=4$; (b) $\mu=2$, and 
(c) $\mu=-1$.
}
\label{I1_I2} 
\end{figure}

\begin{figure}
\begin{center} 
(a)\epsfig{file=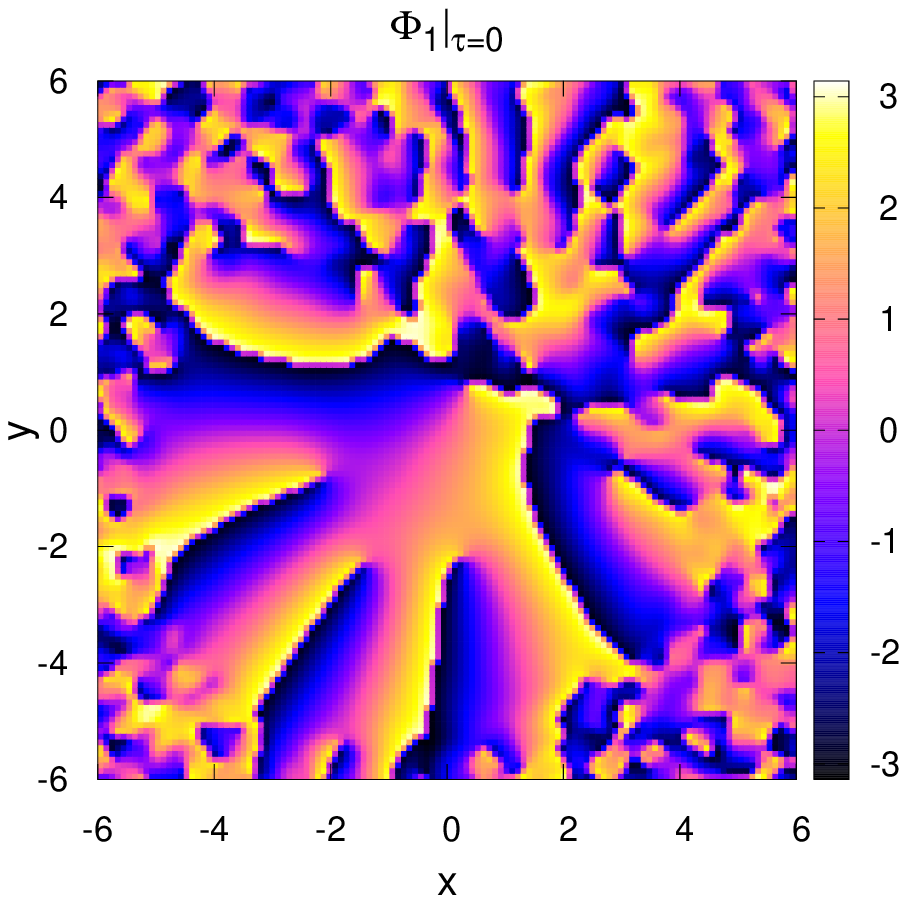, width=75mm}\\
(b)\epsfig{file=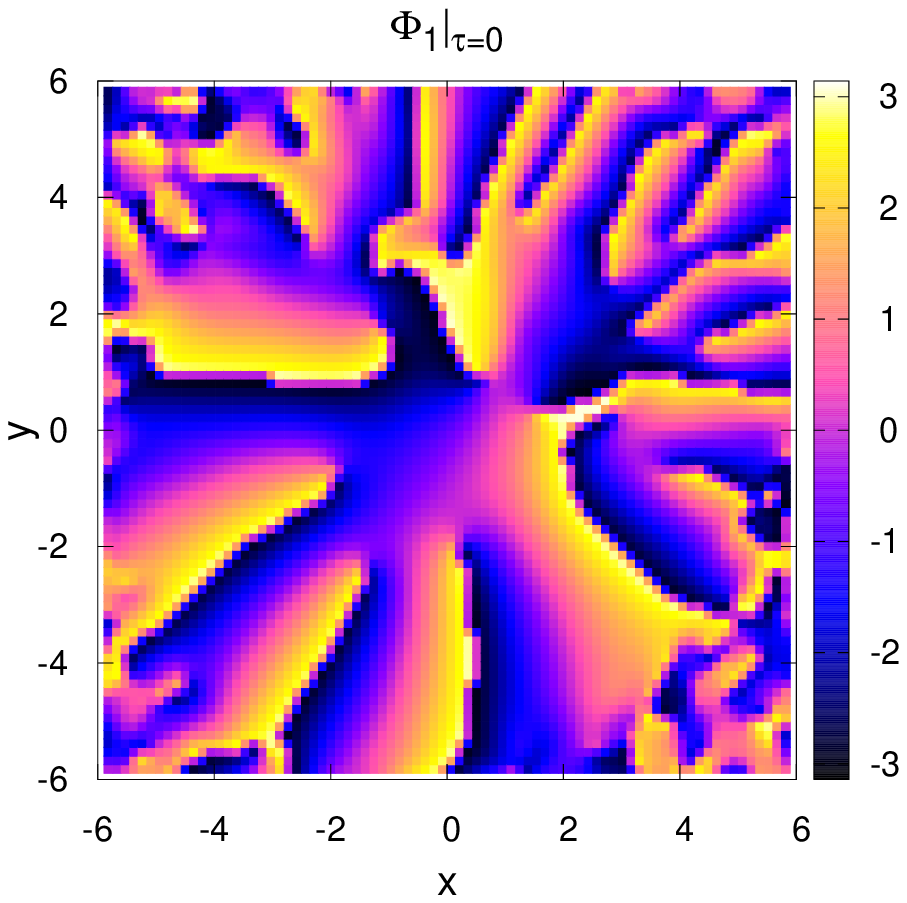, width=75mm}\\
(c)\epsfig{file=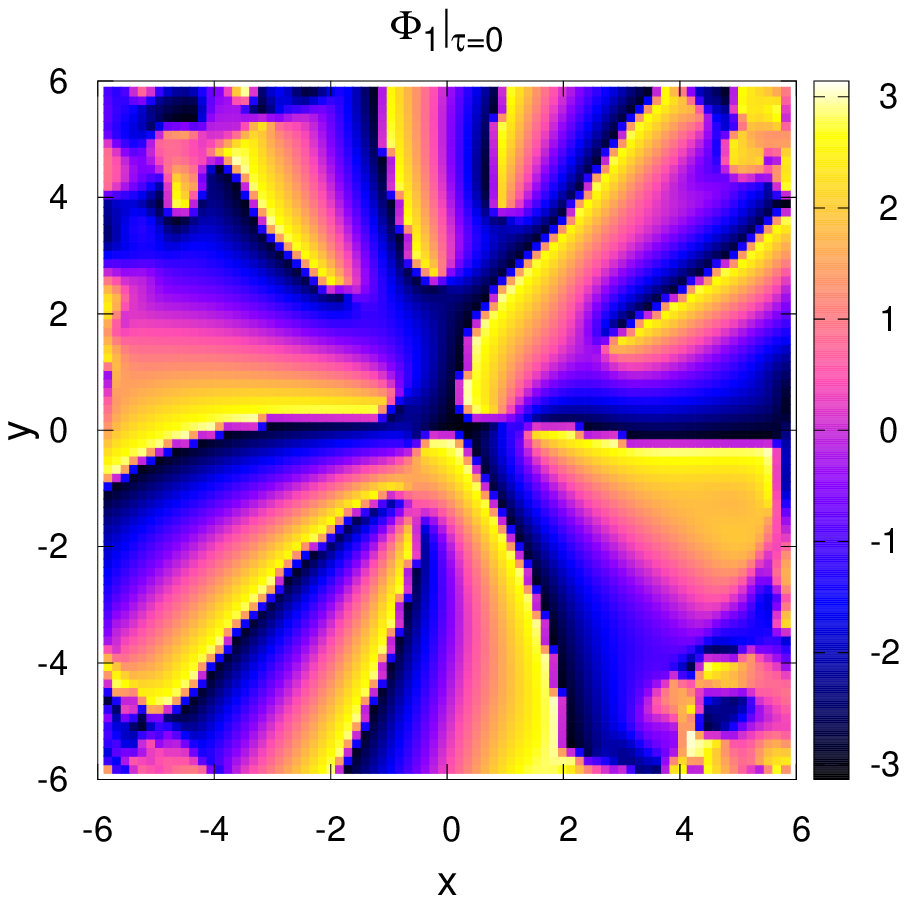, width=75mm}
\end{center}
\caption{Phase $\Phi_1$ of complex envelope $A_1$ for the three configurations
shown in Fig.1. Panels (a), (b), and (c) correspond to those in Fig.1.
}
\label{Phi_1} 
\end{figure}

\section{Results}
    
We have performed two series of numerical experiments. In the first series,
we investigated quasi-2D  configurations for which the dependence on 
variable $\tau$ was weak so that vortex filaments were more or less
parallel to a domain wall. In the second series of experiments, vortices
were oriented as before along the beam, but for $\zeta=0$, two domain
walls (on one period in $\tau$) cut across the beam, thus forming regions
with right- and left-hand polarizations alternating along the beam axis
so that the ends of vortices were attached to the domain walls and strongly
deformed them.

The characteristic examples from the first series are shown in Figs. 1 and 2. 
Since the dynamics remained quasi-2D  on the entire interval of $\zeta$, 
it was sufficient to demonstrate a typical distribution of fields in one 
of cross sections. Since both intensities $I_1$ and $I_2$ simultaneously differ
from nearly zero only over the domain wall thickness, their difference turns
out to be a quite informative quantity. Practically, if it is positive, we
are dealing with the left-polarized region, while when it is negative, we 
have the right-polarized region. It can be seen in Fig.1 that the 
intensities are maximal not at the waveguide center, but on its periphery.
This property is a direct consequence of the action of the “centrifugal force” 
in the rotating coordinate system. The centrifugal force presses “light liquids”
to the waveguide walls. In this respect, Fig.1c corresponding to the negative
value of the chemical potential is especially interesting. It can be seen that
a wide region exists at the middle of the waveguide, where both intensities 
are almost equal to zero. This is exactly a nonlinear variant of the whispering
gallery effect. Figure 2 showing the phase distribution for the first component
corresponding to Fig.1 provides interesting information on the position of
vortices. It can be seen that only a part of vortices is located in the region
with noticeable intensity $I_1$ (patent vortices). These patent vortices avoid
the intensity maxima and, hence, are grouped closer to the center. The other
part (latent vortices) is located in the region where $I_1$ is negligibly small.
In Fig.1c, there are no patent vortices, but there is a dozen of latent vortices 
exactly in the “empty” central region of the waveguide, as follows from Fig.2c.

\begin{figure}
\begin{center} 
(a)\epsfig{file=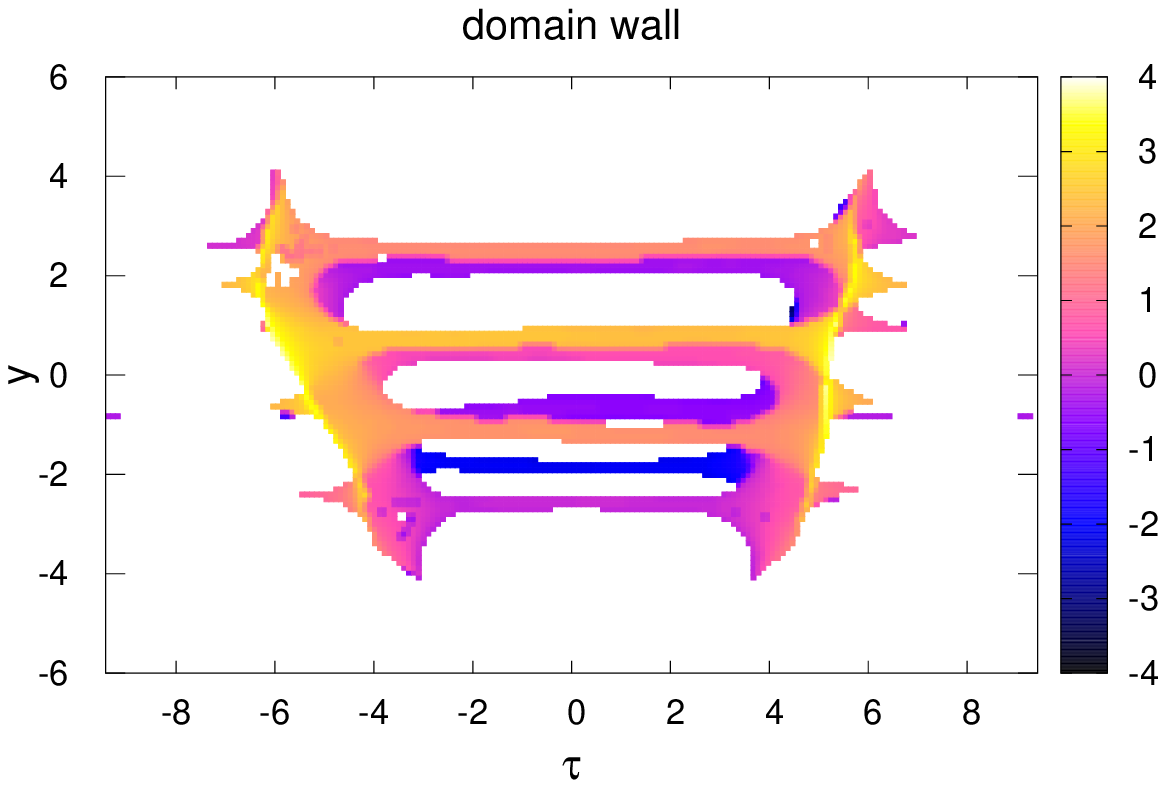, width=80mm}\\
(b)\epsfig{file=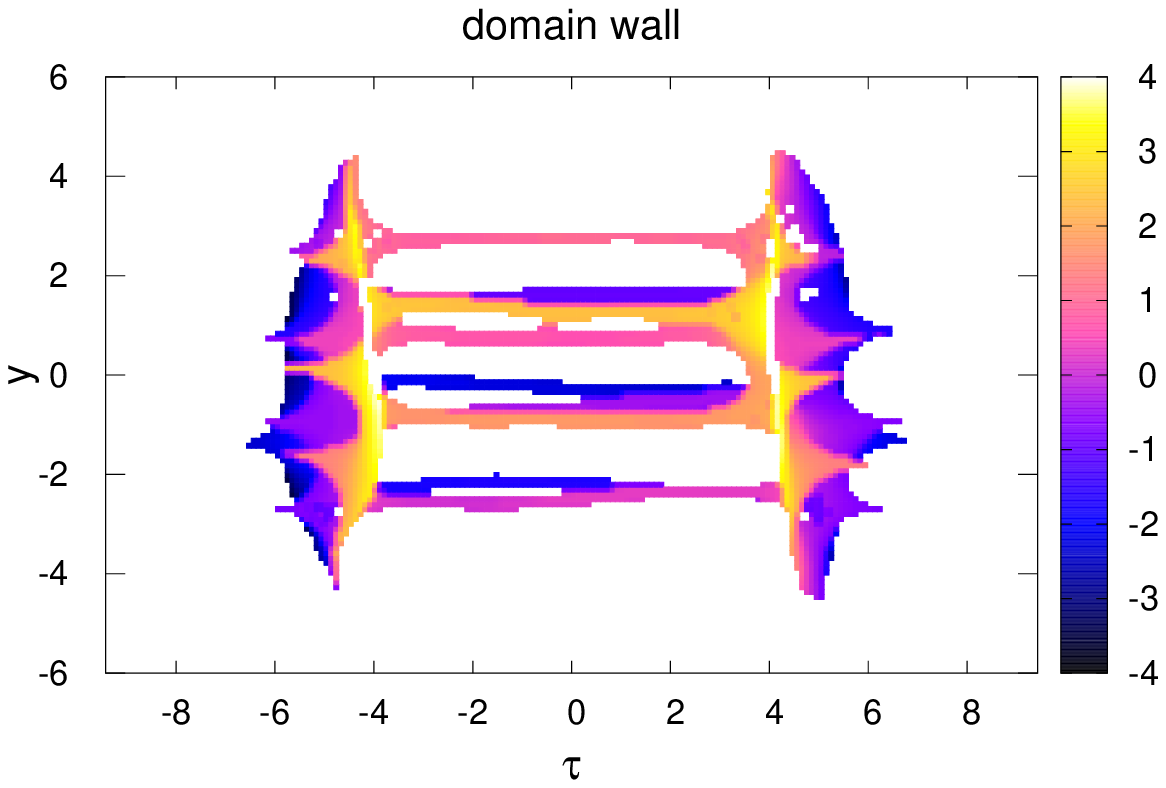, width=80mm}
\end{center}
\caption{ Numerical examples of long-lived transverse domain walls with
attached vortices for $\zeta=500$, $\Omega=0.8$ for (a) $\mu =4$ 
and (b) $\mu =8$. The $x$ coordinate of the points of the numerical grid,
at which $0<(I_1-I_2)<d$, where $d=1.0$ in (a) and $d=1.8$ in (b), 
is marked by the color. This makes it possible to visualize the vortex
cores in the first component as well as a thin layer near the middle 
surface in each domain wall. 
}
\label{walls} 
\end{figure}

As regards the second series of experiments, the essentially 3D evolution
of transverse domain walls with attached vortices proceeds differently for
different values of $\mu$. For small values of $\mu=0$ and $\mu=1$, such
configurations are rapidly “smeared,” and then, after a certain transition 
process, quasi-2D states similar to those described above are formed. 
For larger values of $\mu=4$ and $\mu=8$, although the transverse walls
acquired on the average a tilted orientation depending on $\zeta$, they 
continued their existence in the entire interval of $\zeta$ up to the end
of calculations. Such a change in the modes of behavior is still unclear.
Two examples of 3D structures are shown in Fig.3 (cf. [37]).

It should be noted that for a lower frequency of rotation, the number of 
vortices decreases, and the tendency to the transverse orientation of domain
walls becomes stronger. This tendency conflicts with the aforementioned 
quasi-2D nature of the flows for small values of $\mu$. As a consequence,
a nontrivial intermediate regime becomes possible in a certain range of 
parameters $(\Omega,\mu)$, when one or two vortices and a nonstationary
3D-perturbed domain wall are present; in this regime, the vortices can 
sometimes “hide themselves” in this wall partly or completely. Such a 
dynamics was observed, for example, for $\Omega=0.4$ and $\mu=2$. 
This regime should be investigated in detail.

\section{Conclusions}   
    
Thus, we have traced the theoretical analogy between a coherent optical
wave in a wide helical waveguide and a rotating binary Bose--Einstein 
condensate of cold atoms. Some differences between these two systems are
observed in the form of the transverse confining potential. For atoms, 
a quadratic potential can be obtained more easily, while for an optical 
waveguide, a plane refractive index profile with sharp edges is easier 
for implementation. Accordingly, the effect of the centrifugal force is
stronger in the latter case. This difference leads to different 
quasi-stationary patterns as can be seen from comparison of the figures
in this article with those in [34--37].

This trend in investigations undoubtedly has wide prospects. 
Although only the first steps have been taken in this direction,
nontrivial results have already been obtained and the next goals have
been indicated. First of all, parametric region $(N_1, N_2, \Omega)$ 
should be investigated more systematically. In addition, it is important
from the practical point of view to study transient dynamic processes 
because it is difficult to form quasi-stationary states immediately at 
the waveguide input. Apart from helical waveguides, interesting effects
must also exist for some other dependences of the shape of the cross 
section on coordinate $\zeta$. For example, periodic oscillations of the
cross-sectional area or its other geometrical parameters can be 
responsible for parametric resonances that usually lead in nonlinear 
systems to the formation of certain coherent structures 
(see, for example, [32, 33]).

The adequacy of the simplest model (6) in planning and implementation
of optical experiment requires further serious investigation. In actual
substances, the factors disregarded in this study (e.g., third-order 
dispersion) can play a certain role. It would be interesting in principle
to exceed the limits of the quasi-monochromatic regime and see how the 
wave pattern will change thereby.

\subsection*{Funding}

This study was performed under State assignment no. 0029-2021-0003.

\subsection*{Conflict of interests}

The author declares that he has no conflict of interests.

\vspace{9mm}

\hfill {\it Translated by N. Wadhwa}

\end{document}